\documentclass[twocolumn,showpacs,preprintnumbers,amsmath,amssymb]{revtex4}

\usepackage{graphicx}
\usepackage{dcolumn}
\usepackage{bm}

\newcommand{\tr}{\underset{\vec{s}}{\mathrm{Tr}}}
\newcommand{\tra}[1]{\underset{#1}{\mathrm{Tr}}}
\newcommand{\sign}{{\rm sign}}
\renewcommand{\t}[1]{\tilde{#1}}
\newcommand{\an}[1]{\left\langle #1 \right\rangle}
\def\vec#1{\mbox{\boldmath $#1$}}

\newcommand{\ltsim}{\protect\raisebox{-0.5ex}{$\:\stackrel{\textstyle <}
	{\sim}\:$}}
\newcommand{\gtsim}{\protect\raisebox{-0.5ex}{$\:\stackrel{\textstyle >}
	{\sim}\:$}}

\begin{document}

\preprint{APS/}

\title{Statistical mechanics of lossy data compression \\
       using a non-monotonic perceptron}

\author{Tadaaki Hosaka}
 \email{hosaka@sp.dis.titech.ac.jp}
\author{Yoshiyuki Kabashima}%
 \email{kaba@dis.titech.ac.jp}
\affiliation{Department of Computational Intelligence and
 Systems Science, Tokyo Institute of Technology, Yokohama 2268502, Japan}

\author{Hidetoshi Nishimori}
\email{nishi@stat.phys.titech.ac.jp}
 \affiliation{Department of Physics, Tokyo Institute of Technology,
 Tokyo 1528551, Japan}

\date{\today}


\begin{abstract}
The performance of a lossy data compression scheme for uniformly 
biased Boolean messages is investigated via methods of statistical mechanics. 
Inspired by a formal similarity to the storage capacity problem 
in neural network research, we utilize a perceptron of which 
the transfer function is appropriately designed in order to compress 
and decode the messages. 
Employing the replica method, we analytically show that our 
scheme can achieve the optimal performance known in the framework 
of lossy compression in most cases when the code length becomes infinite.  
The validity of the obtained results is numerically confirmed. 
\end{abstract}

\maketitle


\section{\label{sec:intro}INTRODUCTION}
Recent active research on error-correcting codes (ECC) 
has revealed a great similarity between information theory (IT) and 
statistical mechanics (SM) 
\cite{MacKay,Richardson,Aji,us_EPL,us_PRL,Sourlas_nature,NishimoriWong}. 
As some of these studies have shown that 
methods from SM can be useful in IT, 
it is natural to expect that a similar approach may 
also bring about novel developments in fields other 
than ECC. 

The purpose of the present paper is to offer such an example. 
More specifically, we herein employ methods from SM 
to analyze and develop a scheme of data compression. 
Data compression is generally classified into 
two categories; lossless and lossy compression \cite{bi:Cover}. 
The purpose of lossless compression 
is to reduce the size of messages in information 
representation under the constraint of perfect retrieval. 
The message length in 
the framework of lossy compression can be further reduced by 
allowing a certain amount of 
distortion when the original expression is retrieved. 

The possibility of lossless compression was first pointed out 
by Shannon in 1948 in 
the {\em source coding theorem} \cite{Shannon},
whereas the counterpart of lossy compression, 
termed the {\em rate-distortion theorem}, was presented
in another paper by Shannon more than ten years later \cite{RD}.
Both of these theorems provide the best possible compression 
performance in each framework. However, their proofs
are not constructive and suggest few clues for how to design practical codes. 
After much effort had been made for achieving the 
optimal performance in practical time scales, 
a practical lossless compression code that 
asymptotically saturates the source-coding limit 
was discovered \cite{Jelinek}.
Nevertheless, thus far, regarding lossy compression, no algorithm which 
can be performed in a practical time scale
saturating the optimal performance predicted
by the rate-distortion theory has been found, even for simple 
information sources. Therefore, the quest for better lossy compression 
codes remains one of the important problems in IT
\cite{bi:Cover,fixed, Yamamoto, Murayama_RD}.

Therefore, we focus on designing 
an efficient lossy compression code for a simple 
information source of uniformly biased Boolean sequences.
Constructing a scheme of data compression requires 
implementation of a map from compressed 
data of which the redundancy should be minimized, to the original 
message which is somewhat biased and, therefore, 
seems redundant. 
However, since the summation over the Boolean 
field generally reduces the statistical bias of the data, 
constructing such a map for the 
aforementioned purpose by only linear operations is difficult,
although 
the best performance can be achieved by such linear maps 
in the case of ECC \cite{MacKay,Aji,us_EPL,us_PRL,Sourlas_nature} 
and lossless compression \cite{Murayama}. 
In contrast, producing a biased output  
from an unbiased input is relatively easy when a non-linear map is used. 
Therefore, we will employ a perceptron of which the transfer function 
is optimally designed in order to devise a lossy compression scheme. 

The present paper is organized as follows. In the next section, 
we briefly introduce the framework of lossy data compression,
providing the optimal compression performance 
which is often expressed as the {\em rate-distortion function } 
in the case of the uniformly biased Boolean sequences. 
In section \ref{sec:scheme}, 
we explain how to employ a non-monotonic perceptron to 
compress and decode a given message. 
The ability and limitations of the proposed scheme 
are examined using the replica method in section \ref{sec:app}.~
Due to a specific (mirror) symmetry that we impose on the transfer 
function of the perceptron, one can {\em analytically }
show that the proposed method can saturate the rate-distortion 
function for most choices of parameters when the code 
length becomes infinite.  
The obtained results are numerically validated by means of the 
extrapolation on data from systems of finite size in section 
\ref{sec:extrapo}.~
The final section is devoted 
to summary and discussion.

\section{\label{sec:DC}LOSSY DATA COMPRESSION}
\begin{figure}[h]
	\begin{center}
	\includegraphics{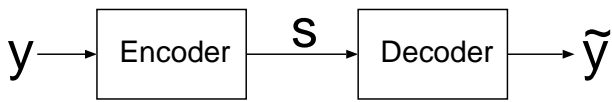}
	\end{center}
    \caption {Encoder and decoder in the framework of lossy compression.
              The retrieved sequence $\t{\vec{y}}$ need not be
              identical to the original sequence $\vec{y}$.}
	\label{fig:coder}
\end{figure}

Let us first provide the framework of lossy data compression. 
In a general scenario, a redundant original 
message of $M$ random variables $\vec{y} = (y^1,y^2,\ldots,y^M)$, 
which we assume here as a Boolean sequence $y^\mu \in \{0,1\}$, is 
compressed into a shorter (Boolean) expression 
$\vec{s} = (s_1,s_2,\ldots,s_N)~(s_i \in \{0,1\}, N<M)$.  
In the decoding phase, the compressed expression $\vec{s}$ is 
mapped to a representative message $\t{\vec{y}} 
= (\t{y}^1,\t{y}^2,\ldots,\t{y}^M)~(\t{y}^\mu \in \{0,1\})$ 
in order to retrieve the original expression (Fig.~\ref{fig:coder}). 

In the source coding theorem, it is shown that 
perfect retrieval $\t{\vec{y}} =\vec{y}$ is possible 
if the compression rate $R={N}/{M}$ is greater than 
the entropy per bit of the message $\vec{y}$ when 
the message lengths $M$ and $N$ become infinite. 
On the other hand, in the framework of lossy data compression, 
the achievable compression rate can be further reduced allowing 
a certain amount of distortion between 
the original and representative messages $\vec{y}$ and 
$\t{\vec{y}}$. 

A measure to evaluate the distortion
is termed the {\em distortion function}, 
which is denoted as $d(\vec{y},\t{\vec{y}})\ge 0$.
Here, we employ the Hamming distance 
\begin{eqnarray}
d(\vec{y},\t{\vec{y}}) = \sum_{\mu=1}^M d(y^\mu,\t{y}^\mu),
\label{eq:Hamming}
\end{eqnarray}
where \begin{eqnarray}
d(y^\mu, \t{y}^\mu) = \left\{ 
	\begin{array}{cc}
        0 & \textrm{if}~~ y^\mu =   \t{y}^\mu \\
		1 & \textrm{if}~~ y^\mu \ne \t{y}^\mu
	\end{array}
\right.  \label{eq:info_hamming}, 
\end{eqnarray}
as is frequently used for Boolean messages. 

Since the original message $\vec{y}$ is assumed to 
be generated randomly, it is natural to evaluate the average 
of Eq.~(\ref{eq:Hamming}). 
This can be performed by averaging $d(\vec{y},\t{\vec{y}})$ with 
respect to the joint probability of $\vec{y}$ 
and $\t{\vec{y}}$ as 
\begin{eqnarray}
\overline{d(\vec{y},\t{\vec{y}})} = \sum_{\vec{y}} 
\sum_{\t{\vec{y}}} P(\vec{y},\t{\vec{y}}) d(\vec{y},\t{\vec{y}}). 
\end{eqnarray}

By allowing the average distortion per bit 
$\overline{d(\vec{y},\t{\vec{y}})}/M$ up to a given permissible 
error level $0 \le D \le 1$, the 
achievable compression rate can be reduced below the entropy per bit. 
This limit $R(D)$ is termed the {\em rate-distortion function}, 
which provides the optimal compression performance in the 
framework of lossy compression. 

The rate-distortion function is formally obtained as a solution 
of a minimization problem with respect to the mutual information 
between $\vec{y}$ and $\t{\vec{y}}$ \cite{bi:Cover}.
Unfortunately, solving the problem is generally difficult and 
analytical expressions of $R(D)$ are not known in most cases. 

The uniformly biased Boolean message in which each component 
is generated independently from an identical distribution 
$P(y^\mu=1)=1-P(y^\mu=0)=p$ is one of the exceptional models 
for which $R(D)$ can be analytically obtained. 
For this simple source, the rate-distortion function becomes
\begin{eqnarray}
	R(D) = H_2(p) - H_2(D), \label{eq:iidR(D)}
\end{eqnarray}
where $H_2(x) =-x \log_2 x - (1-x)\log_2 (1-x)$.

However, it should be addressed here that a practical code 
that saturates this limit has not yet been reported,
even for this simplest model. 
Therefore, in the following, we focus on this information source and 
look for a code that saturates Eq.~(\ref{eq:iidR(D)}) 
examining properties required for good compression performance.


\section{\label{sec:scheme}COMPRESSION BY PERCEPTRON}
In a good compression code for the uniformly biased source, 
it is conjectured that compressed expressions 
$\vec{s}$ should have the following properties:
\begin{itemize}
	\item[\textbf{(I)}]
	In order to minimize loss of information in the original 
        expressions, the entropy per bit in $\vec{s}$ must be maximized. 
	This implies that the components of $\vec{s}$
	are preferably unbiased and uncorrelated. 
	\item[\textbf{(II)}] In order to reduce the distortion, the representative
        message $\tilde{\vec{y}}(\vec{s})$ should be placed close to 
        the typical sequences of the original messages which are biased. 
\end{itemize}

Unfortunately, it is difficult to construct a code that satisfies 
both of the above two requirements utilizing only linear transformations 
over the Boolean field while such maps provide the 
optimal performance in the case of ECC 
\cite{MacKay,Aji,us_EPL,us_PRL,Sourlas_nature} 
and lossless compression \cite{Murayama}.
This is because a linear transformation 
generally reduces statistical bias in messages, which implies that 
the second requirement \textbf{(II)} cannot be realized for unbiased and uncorrelated
compressed expressions $\vec{s}$ that are preferred in 
the first requirement \textbf{(I)}. 

One possible method to design a code that has the above properties 
is to introduce a non-linear transformation. 
A perceptron provides one of the simplest 
schemes for carrying out this task. 

In order to simplify notations, let us replace all the 
Boolean expressions $\{0, 1\}$ with binary ones $\{ 1, -1 \}$. 
By this, we can construct a non-linear map from 
the compressed message $\vec{s}$ to the retrieved sequence $\tilde{\vec{y}}$ utilizing 
a perceptron as 
\begin{eqnarray}
\tilde{y}^\mu= f \left(
\frac{1}{\sqrt{N}} \vec{s} \cdot \vec{x}^\mu \right)~~~
                                    (\mu = 1,2,\ldots,M),
\label{eq:general_decoding}
\end{eqnarray} 
where $\vec{x}^{\mu=1,2,\ldots,M}$ are fixed $N$-dimensional vectors 
to specify the map and $f(\cdot)$ is a transfer function from 
a real number to a binary variable $\t{y}^\mu \in \{1,-1\}$ 
that should be optimally designed. 

Since each component of the original message $\vec{y}$ is 
produced independently, it is preferred to minimize the correlations among 
components of a representative vector $\tilde{\vec{y}}$, 
which intuitively indicates that random selection of $\vec{x}^{\mu}$ 
may provide a good performance. 
Therefore, we hereafter assume that 
vectors $\vec{x}^{\mu=1,2,\ldots,M}$ are independently drawn from 
the $N$-dimensional normal distribution 
$P(\vec{x}) =\left( 2 \pi \right )^{-N/2} \exp \left [
-|\vec{x}|^2/2 \right ]$. 


Based on the non-linear map (\ref{eq:general_decoding}), 
a lossy compression scheme can be defined as follows:
\begin{itemize}
\item {\bf Compression:} 
For a given message $\vec{y}$, find 
a vector $\vec{s}$ that minimizes the distortion 
$d(\vec{y},\tilde{\vec{y}}(\vec{s}))$, 
where $\tilde{\vec{y}}(\vec{s})$ the representative vector
which is generated from $\vec{s}$ by Eq.~(\ref{eq:general_decoding}). 
The obtained $\vec{s}$ is the compressed message. 
\item {\bf Decoding:} 
Given the compressed message $\vec{s}$, 
the representative vector $\tilde{\vec{y}}(\vec{s})$ produced by 
Eq.~(\ref{eq:general_decoding}) provides the approximate 
message for the original message. 
\end{itemize}
Here, we should notice that the formulation 
of the current problem has become somewhat similar 
to that for the storage capacity evaluation of the Ising perceptron
\cite{bi:Krauth,bi:Gardner} regarding $\vec{s}$, $\vec{x}^\mu$ and $y^\mu$ as
``Ising couplings'', ``random input pattern'' and ``random output'', 
respectively. 
Actually, the rate-distortion limit in the current framework
for $D=0$ and $p=1/2$ can be calculated as the inverse of 
the storage capacity of the Ising perceptron, 
$\alpha_c^{-1}$. 

This observation implies that the simplest choice of 
the transfer function $f(u)={\rm sign}(u)$, where
${\rm sign}(u)=1$ for $u \ge 0$ and $-1$ otherwise, does not saturate
the rate-distortion function (\ref{eq:iidR(D)}). This is because 
the well-known storage capacity of the simple Ising perceptron, 
$\alpha_c=M/N \approx 0.83$, means that 
the ``compression limit'' achievable by this monotonic transfer function 
becomes $R_c=N/M=\alpha_c^{-1} \approx 1.20$ and far from 
the value provided by Eq.~(\ref{eq:iidR(D)}) for this parameter choice 
$R(D=0)=H_2(p=1/2)-H_2(D=0)=1$. 
We also examined the performances obtained by the monotonic transfer 
function for biased messages $0<p<1/2$ by introducing an adaptive threshold 
in our previous study \cite{Hosaka} and found that the 
discrepancy from the rate-distortion function becomes large in particular 
for relatively high $R$ while fairly good performance 
is observed for low rate regions. 

Therefore, we have to design a non-trivial function $f(\cdot)$
in order to achieve the rate-distortion limit, which 
may seem hopeless as there are infinitely many degrees 
of freedom to be tuned. 
However, a useful clue exists in the literature of perceptrons, 
which have been investigated extensively during the last decade. 

In the study of neural network, it is widely known that employing 
a non-monotonic transfer function can highly increase the storage
capacity of perceptrons \cite{Monasson}.
In particular, Bex \textit{et al.~}reported that the capacity of 
the Ising perceptron that has a transfer function of 
the reversed-wedge type $f(u)=f_{\rm RW}(u)=\sign(u-k)\sign(u)\sign(u+k)$
can be maximized to $\alpha_c=1$ by setting $k=\sqrt{2 \ln 2}$ \cite{bi:VdB}, 
which implies that the rate-distortion limit $R=1$ is achieved 
for the case of $p=1/2$ and $D=0$ in the current context.
Although not explicitly pointed out in their paper, 
the most significant feature observed for this parameter choice 
is that the Edwards-Anderson (EA) order parameter 
$(1/N) \left |\left \langle \vec{s} \right \rangle \right |^2$
vanishes to zero, where
$\left \langle \cdots \right \rangle$
denotes the average over the posterior distribution given 
$\vec{y}$ and $\vec{x}^{\mu=1,2,\ldots,M}$. 
This implies that the dynamical variable $\vec{s}$ in the 
posterior distribution given $\vec{y}$ and $\vec{x}^{\mu=1,2,\ldots,M}$
is unbiased and, therefore, the entropy 
is maximized, which meets the first requirement \textbf{(I)} addressed above. 
Thus, designing a transfer function $f(u)$ so as to 
make the EA order parameter vanish seems promising 
as the first discipline for constructing a good compression code.

However, the reversed-wedge type transfer function $f_{ \rm RW}(u)$ is not 
fully satisfactory for the present purpose. This is because 
this function cannot produce a biased sequence due to 
the symmetry $f_{\rm RW}(-u)=-f_{\rm RW}(u)$, which means that 
the second requirement \textbf{(II)} provided above would not 
be satisfied for $p \ne 0.5$. 

\begin{figure}
		\begin{center}
		\includegraphics{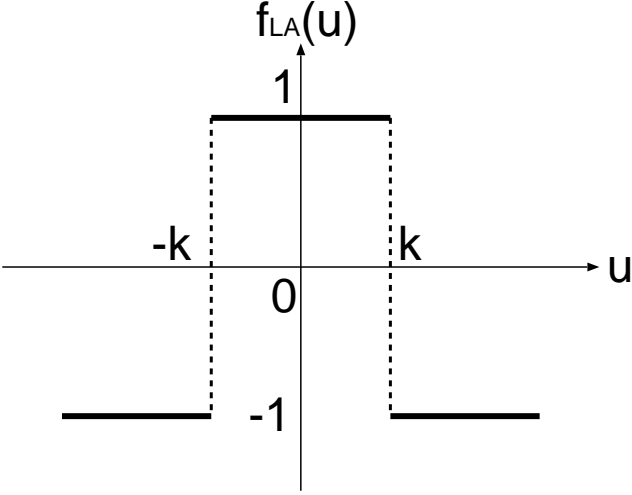}
		\caption {Input-output relation of $f_{\rm LA}(u)$.
                }
		\label{fig:yama1}
		\end{center}
\end{figure}

Hence, another candidate for which 
the EA parameter vanishes and the bias of the output 
can be easily controlled must be found. 
A function that provides these properties was once introduced 
for reducing noise in signal processing, such as 
$f_{\rm LA}(u)=\sign\left (k-|u| \right )$ \cite{bi:Jort,Inoue}
(Fig.~\ref{fig:yama1}). 
Since this locally activated function has mirror 
symmetry $f_{\rm LA}(-u)=f_{\rm LA}(u)$, 
both $\vec{s}$ and $-\vec{s}$ provide identical output for any input,
which means that the EA parameter is likely to be zero. 
Moreover, one can easily control the bias of output sequences 
by adjusting the value of the threshold parameter $k$. 
Therefore, this transfer function looks highly promising as a useful 
building-block for constructing a good compression code. 

In the following two sections, we examine the validity of 
the above speculation, analytically and numerically 
evaluating the performance obtained by the locally activated 
transfer function $f_{\rm LA}(u)$. 

\section{\label{sec:app}ANALYTICAL EVALUATION}
We here analytically evaluate the typical performance of 
the proposed compression scheme using the replica method. 
Our goal is to calculate the minimum permissible 
average distortion $D$ when the compression rate $R = N/M$ is fixed. 
The analysis is similar to that of the storage capacity for perceptrons.

Employing the Ising spin expression, 
the Hamming distortion can be represented as 
\begin{eqnarray}
     d(\vec{y}, \t{ \vec{y} }(\vec{s}) ) = \sum_{\mu = 1}^M
         \left\{ 1 - \Theta_k(u^\mu;y^\mu)  \right\},
		        \label{eq:hamming}
\end{eqnarray}
where 
\begin{eqnarray}
	\Theta_k(u;1)&=& 1-\Theta_k(u;-1) = \left\{
	\begin{array}{ll}
		1, & \mbox{for $|u| \le  k$} \\
		0, & \mbox{otherwise}
	\end{array}
	\right. , \notag \\
	& & \\
	u^\mu & = & \frac{1}{\sqrt{N}}  \vec{s} \cdot \vec{x}^\mu.
\end{eqnarray}
Then, for a given original message $\vec{y}$ and vectors 
$\vec{x}^{\mu(=1,2,\ldots,M)}$, 
the number of dynamical variables $\vec{s}$ which provide 
a fixed Hamming distortion $d(\vec{y}, \t{ \vec{y} }(\vec{s}) ) 
=MD$ $(0 \le D \le 1)$, 
can be expressed as
\begin{eqnarray}
\mathcal{N}(D) = \tr~\delta( MD - d(\vec{y}, \t{ \vec{y} }(\vec{s}))  ). 
\end{eqnarray}

Since $\vec{y}$ and $\vec{x}^\mu$ are randomly generated predetermined 
variables, the quenched average of the entropy per bit over these parameters 
\begin{eqnarray}
 	S(D) = 
\frac{\an{\ln \mathcal{N}(D)}_{ \vec{y}, \vec{x}}  }{N}, 
\end{eqnarray}
to which the raw entropy per bit
$(1/N) \ln \mathcal{N}(D)$ becomes identical for most 
realizations of $\vec{y}$ and $\vec{x}^\mu$, 
is naturally introduced for investigating the typical properties. 
This can be performed by the replica method 
$(1/N)\an{\ln \mathcal{N}(D)}_{ \vec{y}, \vec{x}} 
= \lim_{n \to 0} (1/nN) \ln \an{ \mathcal{N}^n(D)}_{ \vec{y}, \vec{x}}$,
analytically continuing the expressions 
of $\an{ \mathcal{N}^n(D)}_{ \vec{y}, \vec{x}} $
obtained for natural numbers $n$ to non-negative real number $n$ 
\cite{beyond,bi:Nishimori}. 

When $n$ is a natural number, $\mathcal{N}^n(D)$ can be 
expanded to a summation over $n$-replicated systems as
$\mathcal{N}^n(D)=\mathop{\rm Tr}_{\vec{s}^1,\vec{s}^2,\ldots,
\vec{s}^n} \prod_{a=1}^n\delta( MD - d(\vec{y}, \t{ \vec{y}}  (\vec{s}^a
)))$,
where the subscript $a$ denotes a replica index. 
Inserting an identity
\begin{eqnarray}
1&=&\prod_{a>b} \int_{-\infty}^{+\infty} dq_{ab} \delta 
\left (\vec{s}^a \cdot \vec{s}^b - N q_{ab} \right )
\cr
&=& \left (\frac{1}{2\pi i}\right )^{n(n-1)/2}
\int_{-\infty}^{+\infty}
\prod_{a>b} 
dq_{ab} \int_{-i\infty}^{+i\infty}
\prod_{a>b} 
d \hat{q}_{ab} \cr
&& \exp \left [\sum_{a>b}
\hat{q}_{ab} \left (\vec{s}^a \cdot \vec{s}^b - N q_{ab} \right ) \right ]
\end{eqnarray}
into this expression and utilizing the Fourier expression of the delta function
\begin{eqnarray}
&&\delta( MD - d(\vec{y}, \t{ \vec{y}}  (\vec{s}^a ))) \cr
&& =\int_{-i\infty}^{+i \infty} \frac{d\beta_a }{2 \pi i} 
\exp \left [ \beta_a ( MD - d(\vec{y}, \t{ \vec{y}}  (\vec{s}^a ))) \right ], 
\end{eqnarray}
we can calculate the moment $\an{ \mathcal{N}^n(D)}_{ \vec{y}, \vec{x}}$
for natural numbers $n=1,2,3,\ldots$ as 
\begin{widetext}
\begin{eqnarray}
		\an{\mathcal{N}^n(D)}_{ \vec{y}, \vec{x}}
		           & \sim & \int \prod_{a} d\beta_a 
		           \int \prod_{a > b} dq_{a b}
	               \int \prod_{a>b} d\hat{q}_{a b} \notag \\
		       & & \hspace{-2.5cm} 
		       \exp N \left[ R^{-1} \ln 
		       \an{ 
		       \int d\vec{v}~d\vec{u} \exp 
                       \left( -\frac{1}{2} \vec{v}^t Q 
		       \vec{v} + i \vec{v} \cdot \vec{u} \right)
	            \prod_{a=1}^n \left\{
	            e^{-\beta_a} + (1- e^{-\beta_a})
                  \Theta_k \left( u_a;y \right) \right\}
	            }_{y} \right.
                  \notag \\
          & &    \hspace{0.5cm}
           \left.  
	   + \ln \left\{  \tra{\{ s^a \} }~\exp 
           \left( \sum_{a > b} \hat{q}_{a b}
           s^a s^b \right)  
           \right\} 
	   - \sum_{a > b} 
           q_{a b} \hat{q}_{a b}  
           + R^{-1} D \sum_{a=1}^n \beta_a \right], \label{eq:before_sp}
\end{eqnarray}
\end{widetext}
where $Q$ is an $n \times n$ matrix of which elements are 
given by the parameters $\{ q_{ab} \}$ and $\left \langle \cdots \right \rangle_y=
\sum_{y=\pm 1}
\left (p\delta(y-1)+(1-p) \delta(y+1) \right ) \left (\cdots\right )$. 

In the thermodynamic limit $N,M \to \infty$ 
keeping the compression rate $R$ finite, this integral 
can be evaluated via a saddle point problem with respect to 
macroscopic variables $q_{a b}$, $\hat{q}_{a b}$ 
and $\beta_a$.

In order to proceed further, a certain ansatz about 
the symmetry of the replica indices must be assumed. 
We here assume the simplest one, that is, the replica symmetric (RS) ansatz 
\begin{eqnarray}
\beta_a = \beta,~~~q_{ab} 
= q,~~~\hat{q}_{ab} = \hat{q}~~~~~~~~(\forall a>b),
\end{eqnarray}
for which the saddle point expression of Eq.~(\ref{eq:before_sp}) is 
likely to hold for any real number $n$. 
Taking the limit $n \to 0$ of this expression, we obtain 
\begin{widetext}
\begin{eqnarray}
S(D)&=&
 \lim_{n \to 0} \frac{\ln \an{ \mathcal{N}^n(D)}_{ \vec{y}, \vec{x}} }{Nn} 
           \notag \\
  & = & \underset{\beta,q,\hat{q}}{\mathrm{extr}}~~ \left\{ 
 R^{-1} \left[ p \int Dt~ \ln \left\{ e^{-\beta} + (1-e^{-\beta}) 
  			      \left\{ H \left( w_1 \right) 
 	      - H \left(w_2 \right) \right\}    \right\} \right. \right.
  			       \notag \\
 		      & & \hspace{1.5cm} \left. 
 	          + (1-p) \int Dt~ \ln \left\{e^{-\beta} + (1-e^{-\beta}) 
                  \left\{-H \left(w_1 \right) 
                  + H \left(w_2 \right) + 1 
                  \right\} \right\} \right]
                   \notag \\
                 & & \left . \hspace{3.5cm} - \frac{\hat{q}(1-q)}{2} + \int Du~ 
                 \{ \ln (2 \cosh \sqrt{\hat{q}}u) \} 
                           + R^{-1} \beta D \right\},\label{eq:rs_ent}
\end{eqnarray}
\end{widetext}
where $w_1  =  \frac{-k - \sqrt{q}t}{\sqrt{1-q}}$, 
$w_2  =  \frac{k - \sqrt{q}t}{\sqrt{1-q}}$, 
$Dx = \frac{dx}{\sqrt{2 \pi}} \exp(-x^2 / 2)$
 and $H(x) = \int^\infty_x Dt$. 
${\rm extr} \left \{ \cdots \right \}$ denotes the extremization.
Under this RS ansatz, the macroscopic variable $q$ indicates 
the EA order parameter as $q=(1/N)\left 
|\left \langle \vec{s} \right \rangle \right |^2
$. 
The validity of this solution will be examined later. 

\begin{figure}
\begin{center} 
\includegraphics{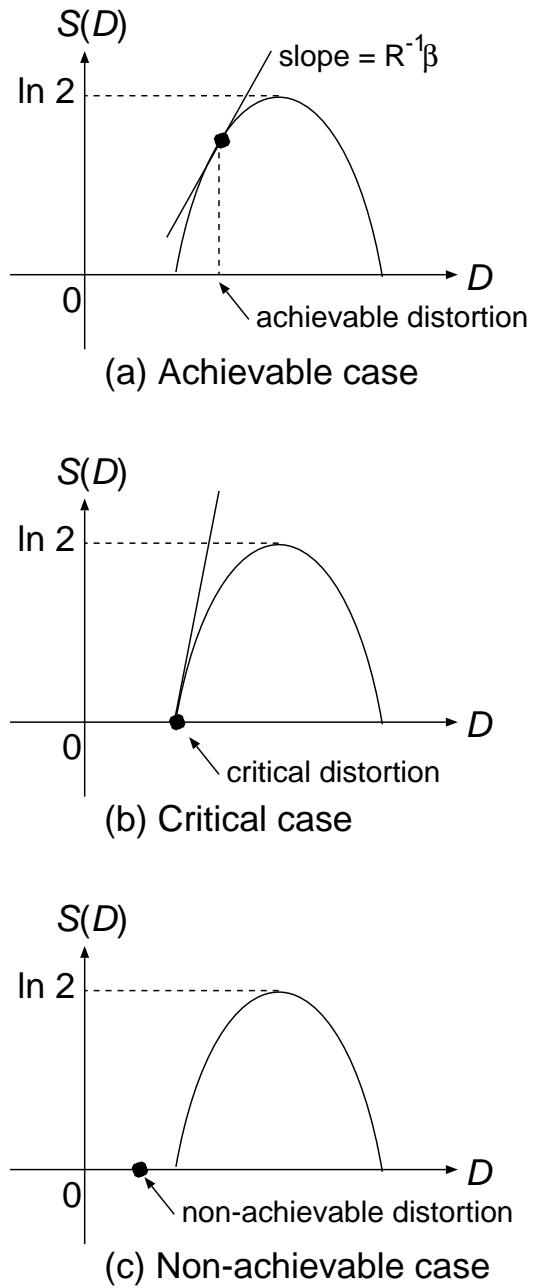}
\caption {Schematic profile of the entropy (per bit) $S(D)$. 
 (a):~For a modest $\beta$, the achieved distortion 
  $D(\beta)$ is such a point that $\partial S \left (D \right )/\partial 
  D =R^{-1}\beta$ holds. 
  This is realized by the random sampling from the 
   canonical distribution $P(\vec{s}|\vec{y},\vec{x}^\mu)\sim 
    \exp[-\beta d(\vec{y},\tilde{\vec{y}}(\vec{s}))]$.  
    (b):~At a critical inverse temperature $\beta=\beta_c$, 
     the entropy for $D(\beta_c)$ which is the minimum 
     distortion vanishes to zero.
    (c):~It is impossible to achieve any distortion 
    which is smaller than $D(\beta_c)$ as $S(D)=0$
    for $D<D(\beta_c)$.}
\label{fig:typical_ent}
\end{center}
\end{figure}

Since the dynamical variable $\vec{s}$ is discrete in the current 
system, the entropy (\ref{eq:rs_ent}) must be non-negative. 
This indicates that the achievable limit for a fixed 
compression rate $R$ and a transfer function $f_{\rm LA}(u)$ which is 
specified by the threshold parameter $k$ 
can be characterized by a transition depicted 
in Fig.~\ref{fig:typical_ent}.

Utilizing the Legendre transformation 
$\beta {F}(\beta)=\mathop{\rm min}_{D}\{
R^{-1} \beta D -{ S}(D)\}$, 
the {\em free energy}
${F}(\beta)$ for a fixed {\em inverse temperature} $\beta$,
which is an external parameter and should be generally distinguished from the
variational variable $\beta$ in Eq.~(\ref{eq:rs_ent}),
can be derived from ${S}(D)$. 
This implies that the distortion $D(\beta)$ that 
minimizes $R^{-1} \beta D -{S}(D)$ and
of which the value is computed from $F(\beta)$ as 
$D(\beta)=\partial(\beta {F}(\beta))/\partial(R^{-1} \beta)$ 
can be achieved by randomly drawing $\vec{s}$ from 
the canonical distribution 
$P(\vec{s}|\vec{y},\vec{x}^\mu)\sim 
\exp[-\beta d(\vec{y},\tilde{\vec{y}}(\vec{s}))]$ 
which is provided by the given $\beta$. 
For a modest $\beta$, the achieved distortion $D(\beta)$
is determined as a point for which 
the slope of ${S}(D)$ becomes identical to $R^{-1}\beta$ 
and ${S}(D)>0$ (Fig.~\ref{fig:typical_ent} (a)).
As $\beta$ becomes higher, $D(\beta)$ moves to the left, which indicates that 
the distortion can be reduced by introducing a lower temperature. 
However, at a critical value $\beta_c$ characterized
by the condition ${S}(D(\beta_c))=0$
(Fig.~\ref{fig:typical_ent} (b)), the number of states that 
achieve $D(\beta_c)$ which is the typical value of 
$\mathop{\rm min}_{\vec{s}}\left 
\{d(\vec{y},\tilde{\vec{y}}(\vec{s})) \right \}$
vanishes to zero. 
Therefore, for $\beta > \beta_c$, 
$D(\beta)$ is fixed to $D(\beta_c)$ and 
the distortion $D<D(\beta_c)$ 
is not achievable (Fig.~\ref{fig:typical_ent} (c)).

The above argument indicates that 
the limit of the achievable distortion $D(\beta_c)$ for 
a given rate $R$ and a threshold parameter $k$ in the 
current scheme can be evaluated from conditions
\begin{eqnarray}
D(\beta)&=&\frac{\partial \left (\beta {F}(\beta)\right )}
{\partial (R^{-1}\beta)}, \label{eq:F2D}\\
{S}(D(\beta))&=&0 \label{eq:S0},
\end{eqnarray}
being parameterized by the inverse temperature $\beta$. 

Due to the mirror symmetry $f_{\rm LA}(-u)=f_{\rm LA}(u)$, $q=\hat{q}=0$ 
becomes the saddle point solution for the extremization problem 
(\ref{eq:rs_ent}) as we speculated in the previous section, 
and no other solution is discovered. 
Inserting $q=\hat{q}=0$ into the right-hand side of Eq.~(\ref{eq:rs_ent})
and employing the Legendre transformation, the free energy 
is obtained as
\begin{eqnarray}
\beta {F}(\beta) &=& -\ln 2 
	-R^{-1} \left[ p \ln \left\{ e^{-\beta} + (1-e^{-\beta}) A_k
  			       \right\} \right.
  			       \notag \\
  			      & & \hspace{-1cm} \left. 
		          + (1-p) \ln \left\{e^{-\beta} + (1-e^{-\beta}) 
                  \left(1 - A_k \right) \right\} \right], \label{eq:0_ent}
\end{eqnarray}
where $A_k=1-2H(k)$, which means that 
Eqs.~(\ref{eq:F2D}) and (\ref{eq:S0}) yield 
\begin{eqnarray}                                  
   D &=& p \frac{ 
	            e^{-\beta} - e^{-\beta} A_k
	            }
	            {
	            e^{-\beta} + (1- e^{-\beta} ) A_k
	            } \notag \\
	   & + &(1-p) \frac{
	            e^{-\beta} - e^{-\beta} ( 1 -A_k )
	            }
	            {
	            e^{-\beta} + (1- e^{-\beta} ) ( 1 - A_k)
	            }, \label{eq:sde_beta_0}                                 
\end{eqnarray}
and 
\begin{eqnarray}
R&=&-\left [p \log_2\left \{ e^{-\beta}+(1-e^{-\beta})A_k \right \} \right . 
\cr
&+&  \left .
(1-p)\log_2\left \{ e^{-\beta}+(1-e^{-\beta})(1-A_k )\right \} \right ] \cr
&-& \frac{\beta}{\ln 2} 
\left [ p \frac{ 
	            e^{-\beta} - e^{-\beta} A_k
	            }
	            {
	            e^{-\beta} + (1- e^{-\beta} ) A_k
	            } \right . \notag \\
	   &   +& \left . (1-p) \frac{
	            e^{-\beta} - e^{-\beta} ( 1 -A_k )
	            }
	            {
	            e^{-\beta} + (1- e^{-\beta} ) ( 1 - A_k)
	            } \right ],\label{eq:R0}
\end{eqnarray}
respectively.

The rate-distortion function $R(D)$ represents the optimal 
performance that can be achieved by appropriately tuning the scheme of 
compression. This means that $R(D)$ can be evaluated as the convex hull
of a region in the $D$-$R$ plane 
defined by Eqs.~(\ref{eq:sde_beta_0}) and (\ref{eq:R0})
by varying the inverse temperature $\beta$ and the threshold parameter $k$ 
(or $A_k$). Minimizing $R$ for a fixed $D$, one can show that 
the relations 
\begin{eqnarray}
	e^{-\beta} & = & \frac{D}{1-D}, \label{eq:op_beta}\\
	e^{-\beta} + ( 1 - e^{-\beta} ) A_k & = & \frac{p}{1-D},
	     \label{eq:op_a1} 
\end{eqnarray}
are satisfied at the convex hull, which offers 
the optimal choice of parameters $\beta$ and $k$ as functions 
of a given permissible distortion $D$ and a bias $p$. 
Plugging these into Eq.~(\ref{eq:R0}), we obtain
\begin{eqnarray}
R &=& R_{\rm RS}(D)=-p \log_2 p - (1-p)\log_2(1-p) \notag \\
     & & \hspace{0.5cm}+ D \log_2 D + (1-D) \log_2 (1-D) \notag \\
& = & H_2(p) - H_2(D),
\label{eq:RS_RD}
\end{eqnarray}
which is identical to the rate-distortion function for 
uniformly biased binary sources (\ref{eq:iidR(D)}).

The results obtained thus far indicate that 
the proposed scheme achieves the rate-distortion limit when 
the threshold parameter $k$ is optimally adjusted. 
However, since the calculation is based on the RS ansatz, 
we must confirm the validity of assuming this specific solution. 
We therefore examined two possible scenarios 
for the breakdown of the RS solution. 

The first scenario is that the local stability against 
the fluctuations for disturbing the replica symmetry is broken, 
which is often termed the Almeida-Thouless (AT) instability \cite{bi:AT},
and can be examined by evaluating the excitation
of the free energy around the RS solution. 
As the current RS solution can be simply expressed as 
$q=\hat{q}=0$, 
the condition for this solution to be stable can be analytically
obtained as
\begin{eqnarray}
	R>R_{\rm AT}(D) = \frac{1}{p(1-p)} \left\{ 
	           \frac{2k(1 -2D)}{\sqrt{2 \pi}} e^{-\frac{k^2}{2} }
	           \right\}^2. \label{eq:AT_sta}
\end{eqnarray}
In most cases, the RS solution satisfies the above condition
and, therefore, does not exhibit the AT instability.
However, we found numerically that for relatively high values of distortion 
$0.336 \ltsim D  < 0.50$, $R_{\rm RS}(D)$ can become 
slightly smaller than $R_{\rm AT}(D)$ for a very narrow 
parameter region, $0.499 \ltsim p \le 0.5$, 
which indicates the necessity of introducing the 
replica symmetry breaking (RSB) solutions. 
This is also supported analytically by the fact that the inequality
$R_{\rm AT}(D)\sim 2.94 \times (p-D)^2 
\ge R_{\rm RS}(D) \sim 2.89 \times
(p-D)^2$ holds for $p=0.5$ in the vicinity of $D \ltsim p$. 
Nevertheless, this instability may not be serious in practice,
because the area of the region $R_{\mathrm{RS}}(D) < R < R_{\mathrm{AT}}(D)$,
where the RS solution becomes unstable, is extremely small, as indicated by Fig.~\ref{fig:extrapo_with_limit}~(a).

The other scenario is the coexistence of an RSB solution
that is thermodynamically dominant while the RS solution 
is locally stable. 
In order to examine this possibility, 
we solved the saddle point problem assuming 
the one-step RSB (1RSB) ansatz in several cases for which 
the RS solution is locally stable. 
However, no 1RSB solution was discovered for $R \ge R_{\rm RS}(D)$. 
Therefore, we concluded that this scenario need not be  
taken into account in the current system.

These insubstantial roles of RSB may seem somewhat surprising 
since significant RSB effects above the storage capacity 
have been reported in the research of perceptrons with continuous
couplings
\cite{bi:Jort,Monasson}. 
However, this may be explained by the fact that,
in most cases, RSB solutions for Ising couplings can be expressed by the RS
solutions adjusting temperature appropriately, 
even if non-monotonic 
transfer functions are used \cite{bi:Krauth, Inoue}. 


\section{\label{sec:extrapo}NUMERICAL VALIDATION}
Although the analysis in the previous section theoretically 
indicates that the proposed scheme is likely to 
exhibit a good compression performance, 
it is still important to confirm it by experiments. 
Therefore, we have performed numerical simulations
implementing the proposed scheme in systems of finite size.

In these experiments, an exhaustive search was performed in order to 
minimize the distortion $d(\vec{y},\tilde{\vec{y}}(\vec{s}))$
so as to compress a given message $\vec{y}$ into $\vec{s}$, 
which implies that implementing the current scheme in 
a large system is difficult.
Therefore, validation was performed by extrapolating 
the numerically obtained data, changing the system size 
from $N=4$ to $N =20$. 

\begin{figure}[t]
  \begin{center}
 \includegraphics[bb = 58 72 753 532, width=8cm,height=7cm]
               {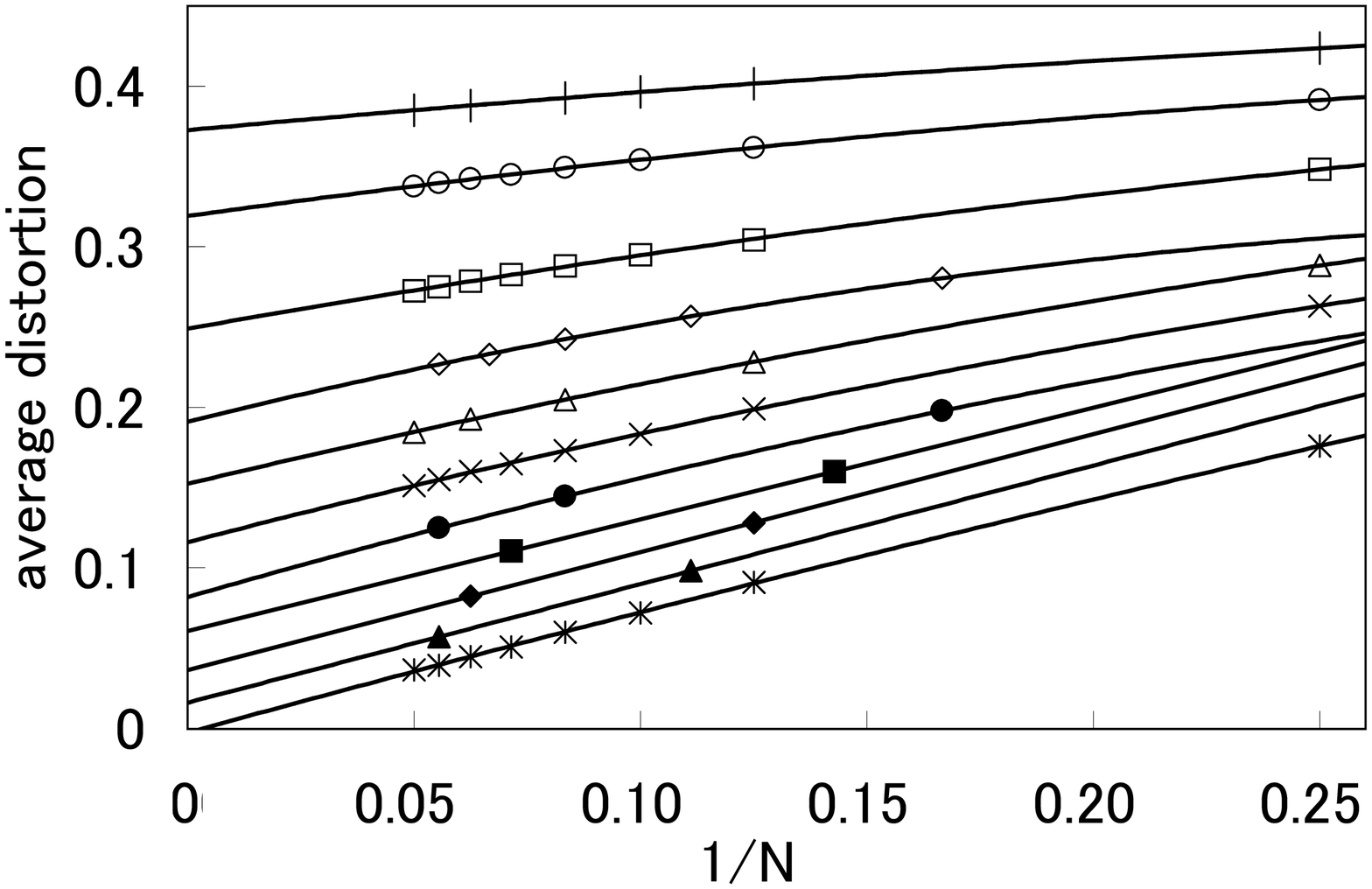} 

(a) $p=0.5$

\vspace{0.7cm}  \includegraphics[bb = 58 72 753 532, width=8cm,height=7cm]
               {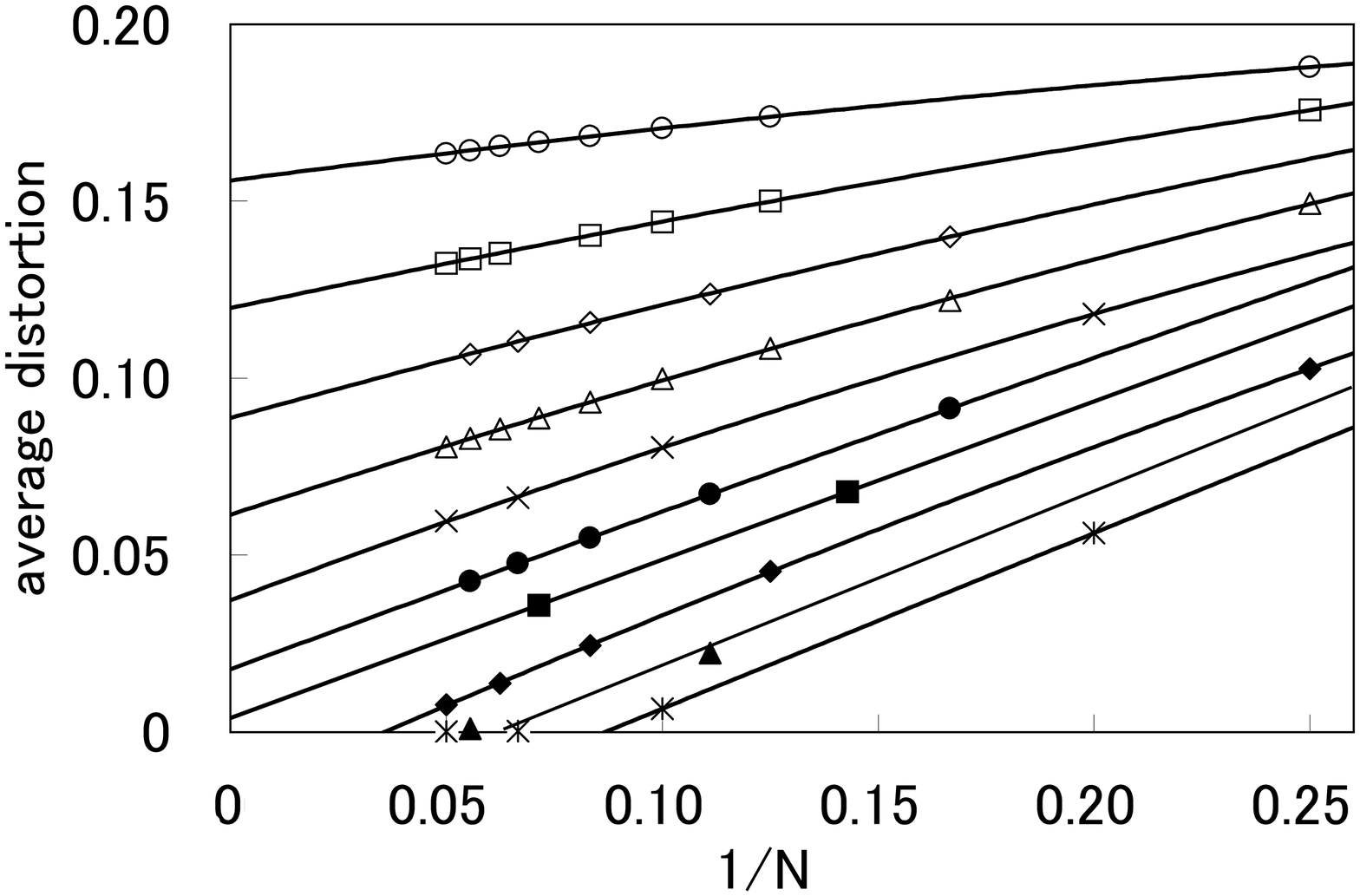} 

(b) $p=0.2$
  \end{center}
  \caption{The averages of the achieved distortions 
are plotted as functions of $1/N$ for 
(a) $p=0.5$ (unbiased) and (b) $p=0.2$ (biased) messages
 changing the compression rate $R$. The plots are obtained from 
 $5000 \sim 10000$ experiments for $N=4 \sim 20$, minimizing 
the distortion $d(\vec{y},\tilde{\vec{y}}(\vec{s}))$ by means of
exhaustive search. 
Each set of plots corresponds to 
$R=$ 0.05 ($p=0.5$ only), 0.1, 0.2,
$\ldots$, 1.0, from the top.
 }
  \label{fig:extrapo}
\end{figure}

Figure~\ref{fig:extrapo} shows the average distortions 
obtained from $5000 \sim 10000$ experiments for 
(a) unbiased ($p=0.5$) and (b) biased ($p=0.2$) messages,
varying the system size $N$ and the compression rate $R(=0.05\sim 1.0)$. 
For each $R$, the threshold parameter $k$ is tuned 
to the value determined using 
Eqs.~(\ref{eq:op_beta}), (\ref{eq:op_a1}) and 
the rate-distortion function $R=R(D)$
in order to optimize the performance. 

%
%
%

\begin{figure}[t]
  \begin{center}
 \includegraphics[width=7.5cm,height=6.5cm]{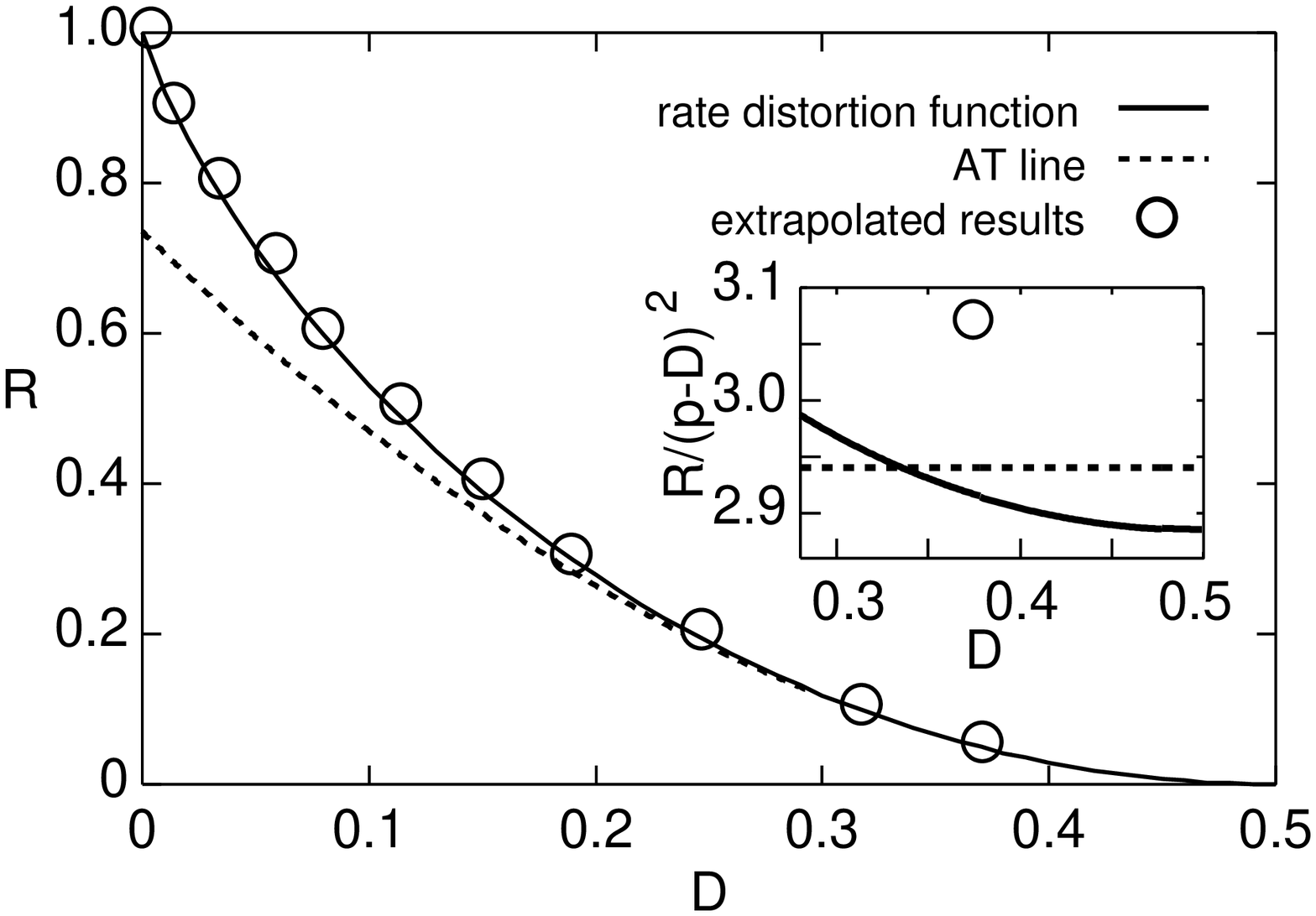} 

(a) $p=0.5$

 \vspace{0.5cm} \includegraphics[width=7.5cm,height=6.5cm]{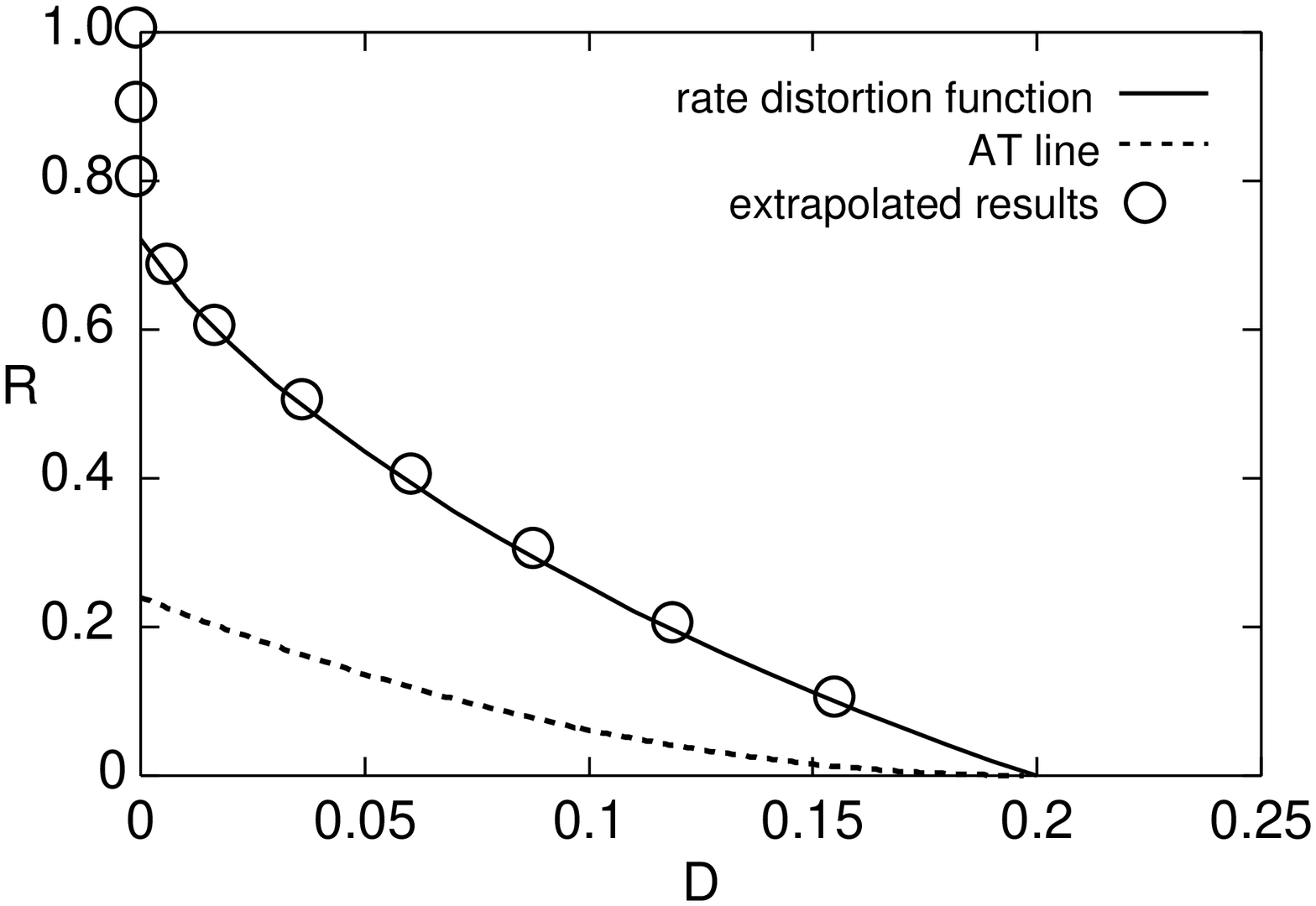} 

(b) $p=0.2$

  \end{center}
  \caption{
The limits of the achievable distortion expected for $N \to \infty$
are plotted versus the code rate $R$ for (a) $p=0.5$ (unbiased) 
and  (b) $p=0.2$ (biased) messages. 
The plots are obtained by extrapolating the numerically 
obtained data for systems of $N=4 \sim 20$ shown in Fig. 
\ref{fig:extrapo}.  The full and dashed curves represent 
the rate-distortion functions and the AT lines, respectively. 
Although the AT stability is broken for $D\gtsim 0.336$ for $p=0.5$ 
(inset of (a)), the numerical data is highly consistent with 
the RS solution which corresponds to the rate-distortion function. }
  \label{fig:extrapo_with_limit}
\end{figure}

These data indicate that the finite size effect is relatively 
large in the present system, which is similar to the case of the storage 
capacity problem \cite{bi:Opper}, and do not necessarily seem
consistent with the theoretical prediction obtained in 
the previous section. 
However, the extrapolated values obtained from 
the quadratic fitting with respect to $1/N$ are
highly consistent with curves of the rate-distortion function 
(Fig.~\ref{fig:extrapo_with_limit} (a) and (b)), including 
one point in the region where the AT stability is broken (inset of 
Fig.~\ref{fig:extrapo_with_limit}(a)), 
which strongly supports the validity and efficacy 
of our calculation based on the RS ansatz.

%
%
%
%

\section{\label{sec:summary}SUMMARY AND DISCUSSION}
We have investigated a lossy data compression 
scheme of uniformly biased Boolean messages employing a perceptron of 
which the transfer function is non-monotonic. 
Designing the transfer function based on the properties required for 
good compression codes, we have constructed 
a scheme that saturates the rate-distortion function that 
represents the optimal performance in the framework 
of lossy compression in most cases. 

It is known that a non-monotonic single layer perceptron can be 
regarded as equivalent to certain types of multi-layered networks, 
as in the case of parity and committee machines. 
Although tuning the input-output relation in 
multi-layered networks would be more complicated, 
employing such devices might be useful in practice 
because several heuristic algorithms that could be 
used for encoding in the present context 
have been proposed and investigated \cite{Mitchison,Nokura}.

In real world problems, the redundancy of information sources
is not necessarily represented as a uniform bias; but rather is
often given as non-trivial correlations among components of a message. 
Although it is just unfortunate that the direct employment of 
the current method may not show a good performance in such cases, 
the locally activated transfer function $f_{\rm LA}(u)$ that 
we have introduced herein could serve as a useful building-block 
to be used in conjunction with a set of connection 
vectors $\vec{x}^{\mu=1,2,\ldots,M}$ that are appropriately correlated 
for approximately expressing the given information source,
because by using this function, 
we can easily control the input-output relation suppressing
the bias of the compressed message to zero, no matter how 
the redundancy is represented. 


Finally, although we have confirmed that our method exhibits 
a good performance when executed optimally in a large
system, the computational cost for compressing a message 
may render the proposed method impractical.
One promising approach for resolving this difficulty 
is to employ efficient approximation 
algorithms such as various methods of the Monte Carlo 
sampling \cite{bi:Sampling} and of the mean field approximation
\cite{bi:Advanced_Mean_Field_Methods}. 
Another possibility is to reduce the finite size 
effect by further tuning the profile of 
the transfer function. 
Investigation of these subjects is currently under way. 

Grants-in-Aid 
Nos.~13780208, 14084206 (YK) and 12640369 (HN) from MEXT are gratefully 
acknowledged. 
TH would like to thank T. Murayama for informing his preprint and valuable
comments, and
YK would like to acknowledge D. Saad, H. Yamamoto and Y. Matsunaga for 
their useful discussions.


\end{document}